\documentclass{PoS}

\usepackage{latexsym}
\usepackage{subfigure}
\usepackage{multicol}
\usepackage{multirow}
\usepackage{mathrsfs}
\usepackage{verbatim}
\usepackage{amsmath,amsthm, amssymb}

\title{The CLS 2+1 flavor simulations} 
\ShortTitle{The CLS 2+1 flavor simulations}
\author{
	\speaker{Piotr Korcyl} 	
\hfill{\footnotesize{\it DESY 14-224}}\\
        John von Neumann Institute for Computing (NIC), DESY, Platanenallee 6, 15738 Zeuthen, Germany\\
        E-mail: \email{piotr.korcyl@desy.de}}

\definecolor{pink}{rgb}{1.0, 0.0, 1.0}

\abstract{
We report on the status of large volume simulations with 2+1 dynamical
fermions which are being performed by the CLS initiative. 
The algorithmic
details include: open boundary conditions, twisted mass reweighting and RHMC, whereas the main feature of the simulation strategy is
the approach to the physical point along a trajectory of constant trace of the mass matrix.
We comment on the practical side of the above issues using as examples some of the
newly generated ensembles, which presently cover lattice spacings between 0.05 fm and 0.11 fm and
pion masses between 150 MeV and 415 MeV.
}

\FullConference{The 32nd International Symposium on Lattice Field Theory\\
                 23-28 June, 2014\\
                 Columbia University New York, NY}

\begin{document}

In order to suppress or eliminate systematic errors associated with the continuum and chiral extrapolations, 
modern lattice QCD simulations try to explore the parameter space of $a \sim 0.05$ fm and physical pion mass.
This is possible not only because of the increase in the available computer resources but mainly because of 
the progress in our understanding of numerical algorithms. New algorithmic developments are aimed at alleviating 
the overall scaling with $a$ and $m_{\pi}$ and try to deal with three phenomena which mainly contribute to the total 
cost of a simulation. These are: the autocorrelations which grow as the continuum limit is approached, 
partially because of the topology freezing; the accidental zero-modes 
of the Dirac operator which render the simulation unstable and finally, the 
growth of the condition number of the Dirac operator as the light quark masses are lowered. The algorithmic 
improvements trying to circumvent the above issues are the following:
\begin{itemize}
\item open boundary conditions in time \cite{luscher_schaefer_11} and periodic boundary conditions 
in the space directions, are supposed to let 
the topological charge flow in and out of the simulated volume and thus decrease the autocorrelation times,
\item twisted mass infrared regulator \cite{luscher_palombi_08} which, by moving the spectrum of the Dirac 
operator off the real axis, decreases the effects of the accidental zero-modes,
\item deflated solver \cite{luscher_04} which, 
makes the cost of the inversion largely independent of the quark mass.
\end{itemize}
The three mentioned ingredients were implemented in the open-source package \verb[openQCD[ \cite{openQCD} and used in  
large volume simulations with 2+1 flavors of non-perturbatively improved Wilson fermions \cite{2+1}. 
The simulations were made by a joint effort of the CLS \cite{cls} collaboration with computer time that was granted 
through two PRACE projects (at LRZ's SuperMuc in Munich (Germany) and CINECA's Fermi in Bologna (Italy)) and three national 
projects (Gauss-Center in J\"ulich (Germany),  NIC in J\"ulich (Germany), CSCS in Lugano (Switzerland)).

In the following we present some of the first impressions gathered during these simulations. We start by briefly describing the 
algorithmic details, then we discuss the strategy to reach the physical point, and finally we concentrate on several observables
such as $t_0$, the topological charge evaluated at positive Wilson flow time or the twisted mass and rational approximation 
reweighting factors evaluated on several of our new ensembles (for more details see Re.\cite{2+1}).

\section{General setup: openQCD-1.2}

The described simulations are being made using the open-source package \verb[openQCD[ version 1.2 \cite{openQCD}. 
The package implements the Hybrid 
Monte Carlo algorithm along with several solvers enabling simulations of dynamical fermions. The gauge action 
used is the tree-level Symanzik improved action ($c_0 = \frac{5}{3}$, $c_1 = - \frac{1}{12}$), while the 
fermion action is discretized with the  Wilson $\mathcal{O}(a)$-improved action with $c_{\textrm{SW}}$ determined 
non-perturbatively \cite{bulava_schaefer_13}. Open boundary conditions for the gauge fields
are implemented following \cite{luscher_10, luscher_schaefer_11}
\begin{equation}
F_{0k}(x)\big|_{x_0=0} = F_{0k}(x)\big|_{x_0=T} = 0, \qquad k=1,2,3, \nonumber
\end{equation}
for the gauge fields and 
\begin{align}
P_+\psi(x)\big|_{x_0=0} &= P_- \psi(x)\big|_{x_0=T} = 0, \qquad P_{\pm} = \frac{1}{2} (1 \pm \gamma_0) \nonumber \\
\bar{\psi}(x)P_-\big|_{x_0=0} &= \bar{\psi}(x)P_+\big|_{x_0=T} = 0, \nonumber
\end{align}
for the fermion fields. The twisted mass infrared regulator is introduced for the two degenerate light fermions \cite{luscher_palombi_08}
by replacing the 
original determinant $\det(D^{\dagger} D)$ by $\frac{\det(D^{\dagger} D + \mu_0^2)^2}{\det(D^{\dagger} D + 2 \mu_0^2)}$ in the simulation
algorithm and reweighted to the proper distribution at the analysis stage. 
Furthermore, the later determinant is split 
into typically $n \sim 5-6$ parts and each part is associated to an independent pseudofermion field,
\begin{equation}
\frac{\det(D^{\dagger} D + \mu_0^2)^2}{\det(D^{\dagger} D + 2 \mu_0^2)} = \det(D^{\dagger} D + \mu_n^2) 
\frac{\det(D^{\dagger} D + \mu_0^2)}{\det(D^{\dagger} D + 2 \mu_0^2)} \prod_{k=0}^{n-1}
\frac{\det(D^{\dagger} D + \mu_k^2)}{\det(D^{\dagger} D + \mu_{k+1}^2)} 
\end{equation}
The dynamical strange quark is implemented with the RHMC algorithm with reweighting \cite{luscher_schaefer_13}. 
All contributions to the HMC forces are integrated along the trajectory with a set of nested MD integrators featuring a 
 3 level hierarchy ($\sim 10$ steps on the coarsest level, one step on finer levels) where 2$^{nd}$ (level 2) 
and 4$^{th}$ (levels 0 and 1) order Omelyan integrators are used.

\section{Strategy: our way to the physical point}

\begin{figure}
\begin{center}
\includegraphics[width=0.49\textwidth]{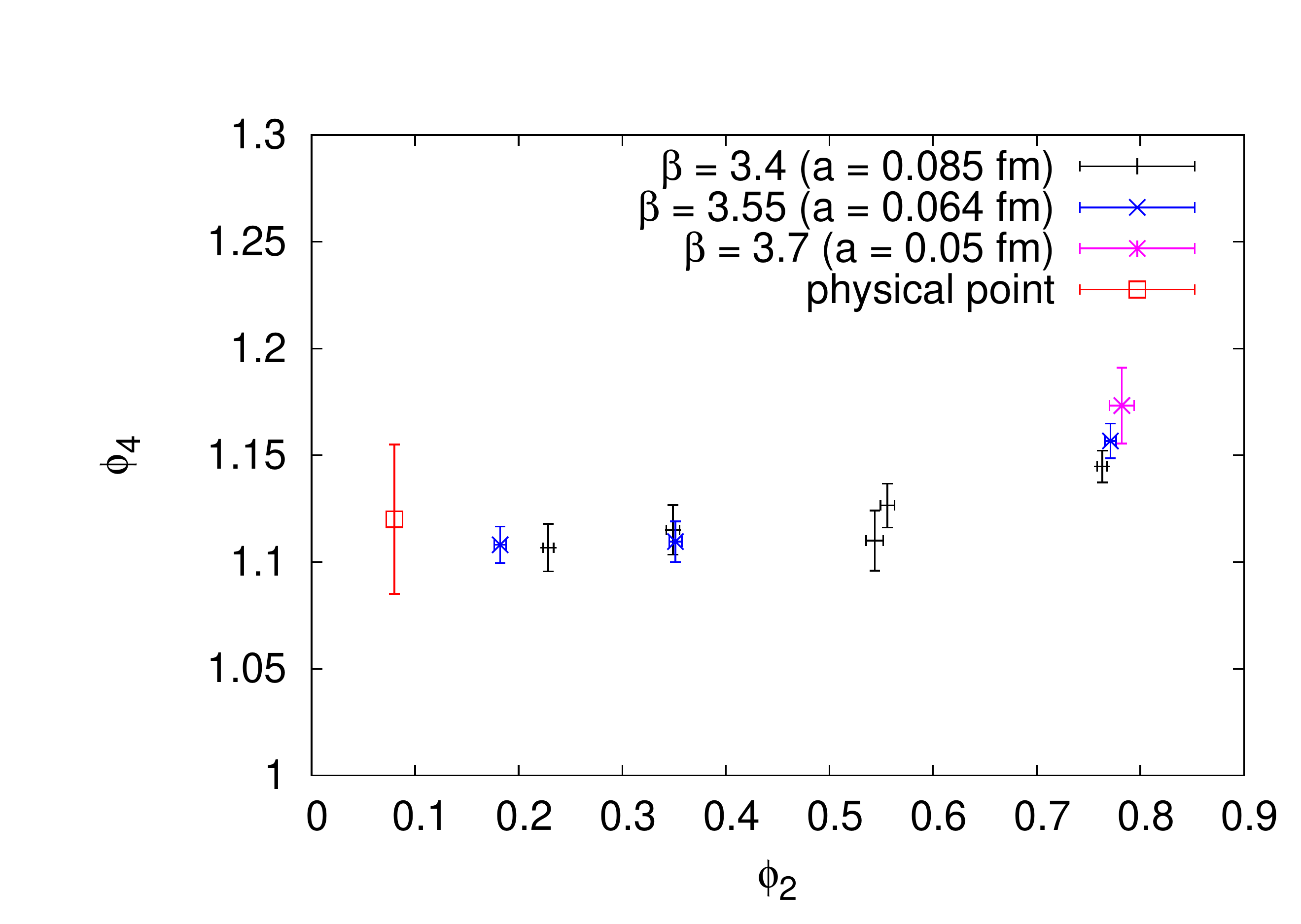}
\caption{Tuning to the trajectory of constant trace of the mass matrix. \label{fig. trajectory}}
\end{center}
\end{figure}

As in all such projects, one has to choose the starting point in parameter space and define the trajectory along which
the continuum and chiral limits are taken. To do so, we define two dimensionless quantities which are used to perform the matching:
\begin{align}
\phi_2(\beta, \kappa_u, \kappa_s) &= 8 t_0 m_{\pi}^2 \propto m_{ud} + \mathcal{O}(m_{ud}^2) \\
\phi_4(\beta, \kappa_u, \kappa_s) &= 8 t_0 (m_K^2 + \frac{1}{2} m_{\pi}^2) \propto \textrm{tr} M + \mathcal{O}(\textrm{tr} M ^2)
\end{align}
Our trajectory is defined by keeping the trace of the mass matrix constant, namely
\begin{equation}
\sum_{i=u,d,s} \frac{1}{\kappa_i} = \textrm{const} \Leftrightarrow \textrm{tr} M_{\textrm{R}} = \textrm{const} + \mathcal{O}(a) 
\end{equation}
In particular, at the symmetric point $\kappa_u = \kappa_d = \kappa_s$ one is left with only one free parameter and therefore
the matching for different lattice spacings can be done relatively easily. Figure \ref{fig. trajectory} presents the results of the 
performed matching. The three points on the right (with $\phi_2 \approx 0.75$) come from the three matched ensembles at the symmetric point, 
whereas the remaining data points have different values of $\phi_2$ but should lie along the line of $\phi_4 \approx \textrm{const}$. 
The estimate of the physical point uses the values of $m_{\pi}$ and $m_{K}$ taken from PDG and $\sqrt{t_0} = 0.1465 \ \textrm{fm}$ 
from \cite{borsanyi_12}.

\begin{table}
\begin{center}
\begin{tabular}{|cc|cccc|c}
\hline
& & 0.085 fm & 0.064 fm & 0.05 fm & a [fm]\\
$m_{K}$ & $m_{\pi}$ & 3.4 & 3.55 & 3.7 & $\beta$\\
\cline{1-6}
{\color{red}415 MeV} & {\color{red}415 MeV} & {\color{red}32$^3\times$96} & {\color{red}32$^3\times$96} & {\color{red}48$^3\times$128} & \\
440 MeV & 350 MeV & 32$^3\times$96 &    &    & \\
470 MeV & 280 MeV & 32$^3\times$96 & 48$^3\times$128 & {\color{blue}64$^3\times$192} & \\
480 MeV & 220 MeV & 48$^3\times$96 & 64$^3\times$128 &    & \\
490 MeV & 150 MeV & {\color{blue}64$^3\times$128} &    &    & \\
\cline{1-6}
\end{tabular}
\end{center}
\caption{Overview table of lattice spacings and pion masses. The red entries correspond to the ensembles at the symmetric point that were
used in the matching between different lattice spacings. The blue entries correspond to the ensembles being still 
generated. \label{tab. ensembles}}
\end{table}

\section{First impressions}

During the granted computer time 10 ensembles were generated with MC chains of length of at least 50 $\tau^{exp}$, whose
details are summarized in table \ref{tab. ensembles} (for more information see Ref.\cite{2+1}). Simultaneously with the 
generation of configurations several
quantities were monitored, such as the topological charge and the action density at positive Wilson flow time and the two 
reweighting factors. We now discuss them in more detail.


\subsection{Reweighting factors}

Before an average value of any physical observable is computed, one has to measure the twisted mass and rational 
reweigthing factors, used to be denoted by $W_0$ and $W_1$, respectively. They are given by
\begin{equation}
W_0 = \frac{\det(D^{\dagger}D) \det(D^{\dagger} D + 2 \mu^2)}{\det(D^{\dagger} D + \mu^2)^2},  \qquad W_1 = \det(D_s R) ,
\end{equation}
with $Z = D^{\dagger}_s D_s R^2 - 1$, where $\det D_s = W_1 \det R^{-1}$ and $R$ is the Zolotarev rational approximation. 
We stress that the inaccuracy of the rational approximation is dealt with in these simulations by reweighting instead of 
an additional Metropolis step as is done typically. 

\begin{figure}
\begin{center}
\includegraphics[width=0.325\textwidth]{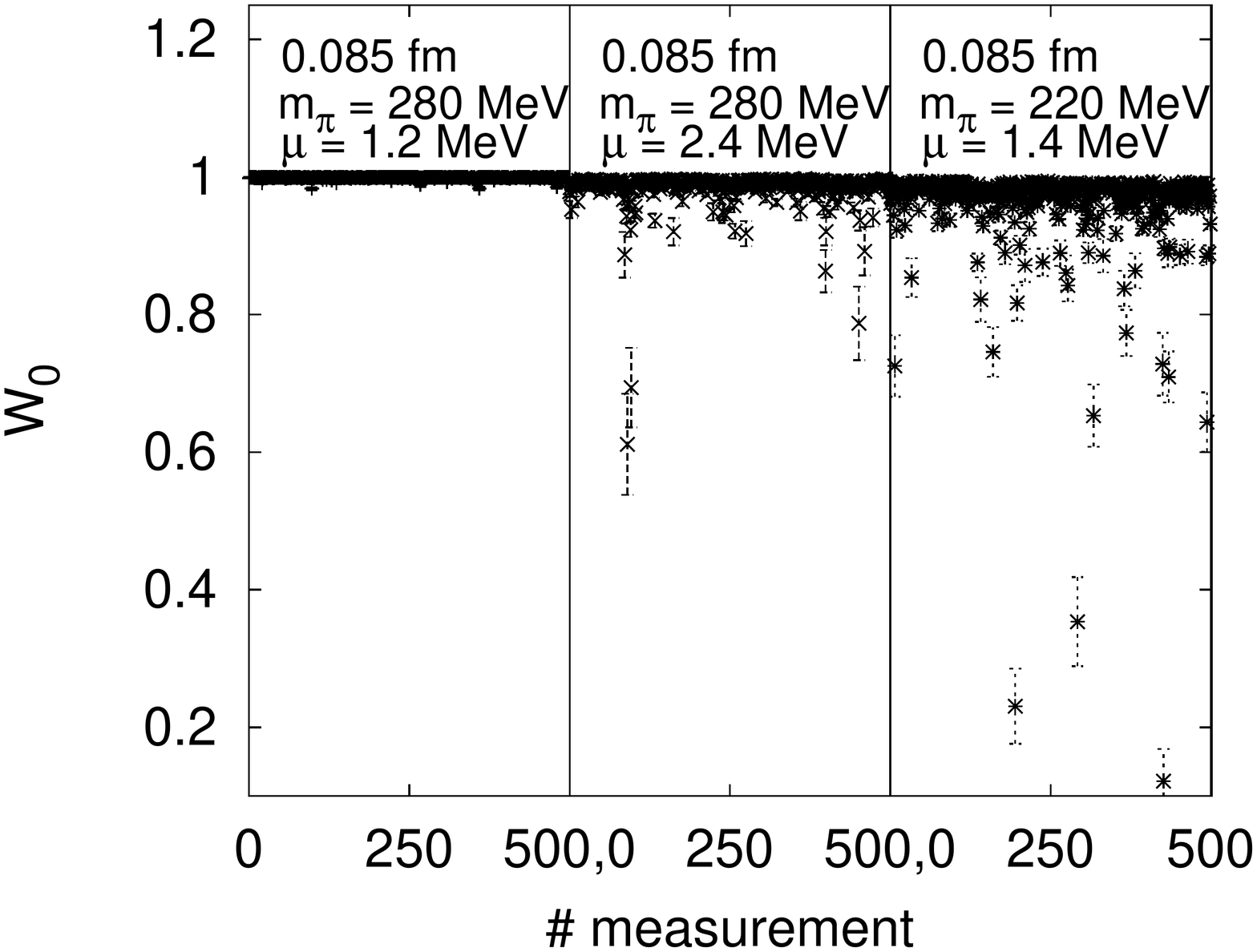}
\includegraphics[width=0.325\textwidth]{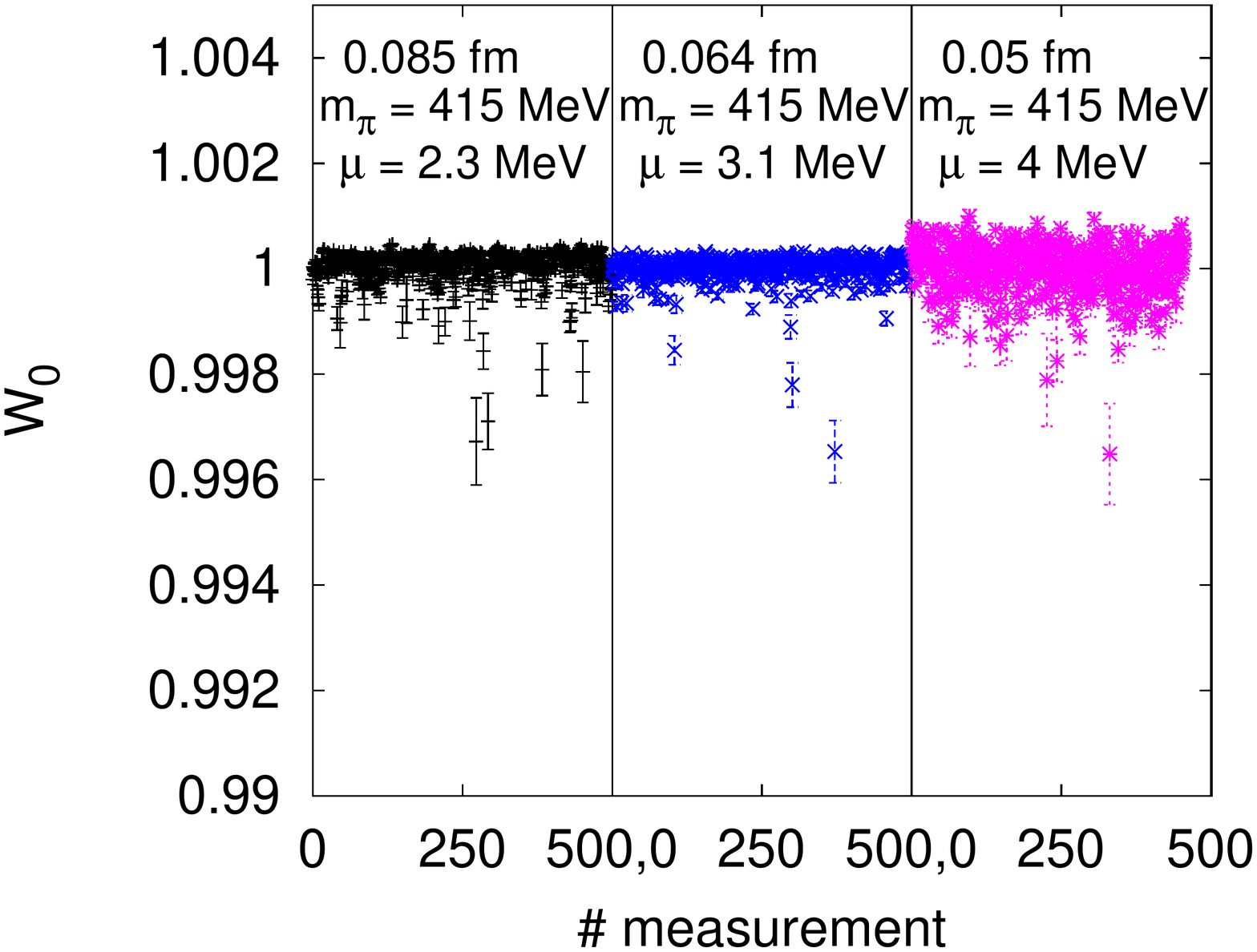}
\includegraphics[width=0.325\textwidth]{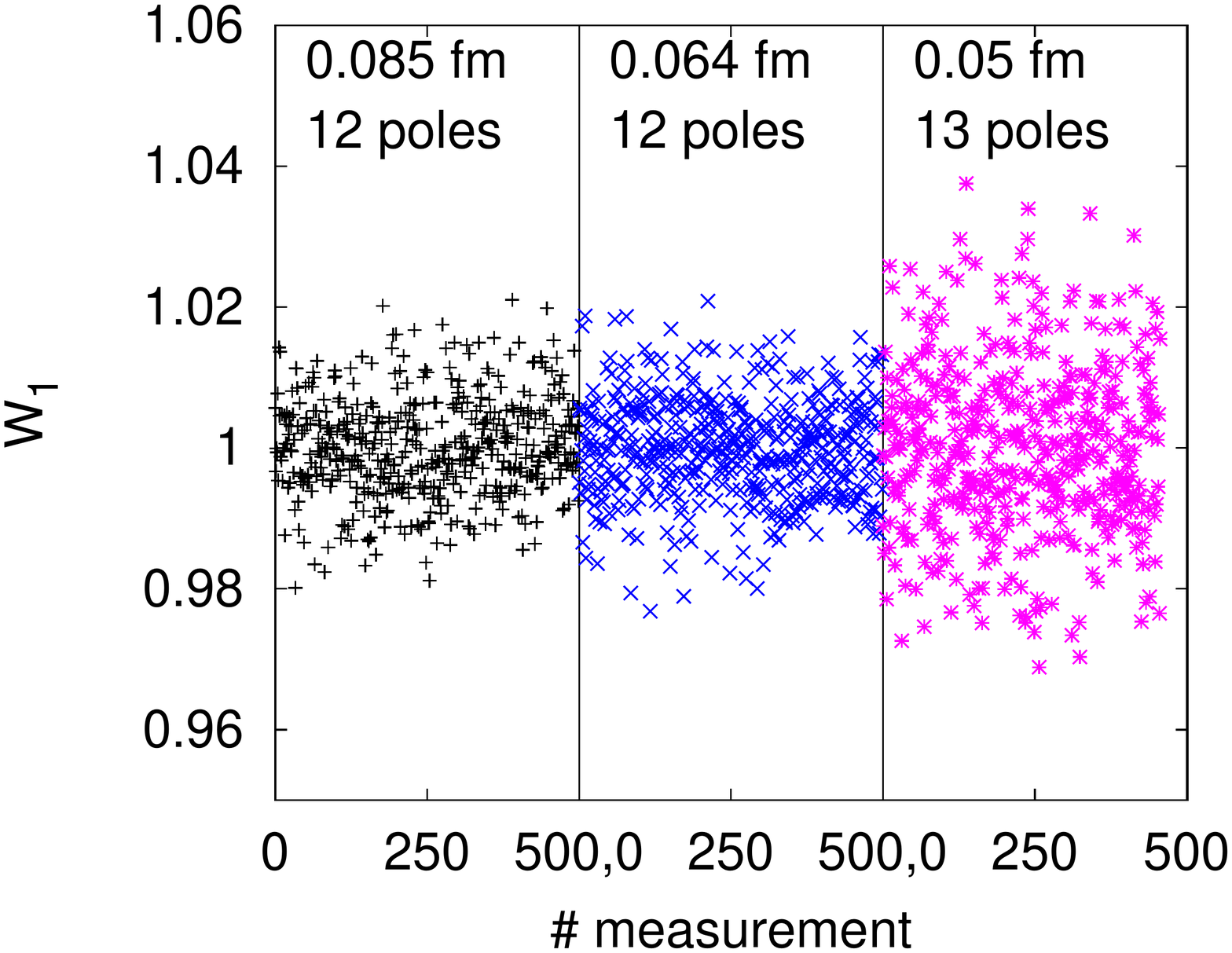}
\caption{Histories of the reweigthing factors. On the left plot we compare $W_0$ of the three ensembles having the same lattice
spacing but different $m_{\pi}$ or $\mu$, the plot in the middle shows the comparison of $W_0$ evaluated
at the symmetric ensembles at different lattice spacings, whereas the plot on the right shows 
the histories of the rational approximation reweighting factor at the symmetric point. Plotted data comes from
ensembles: left panel: H105r005, H105r001, C101r014; middle panel: H101r001, H200r000, N300r002; 
right panel: H101r001, H200r000, N300r002, see Ref.\cite{2+1}. \label{fig. twisted rw}}
\end{center}
\end{figure}

The left and middle panels of figure \ref{fig. twisted rw} show samples of MC histories of the twisted mass reweigthing factor for different ensembles. 
Their values fluctuate close to $1$ as expected. From the algorithmic point of view there is nothing wrong with 
having $W_0$ close to or equal to zero, however such small reweighting factors may lead to a decrease of the 
precision of the final average \cite{mattia}, therefore they have to be monitored during the entire production phase.

As far as the rational reweighting factors are concerned, the right panel of figure \ref{fig. twisted rw} shows extracts of MC histories 
of $W_1$ for the three ensembles at the symmetric point at three different 
lattice spacings. We see that the fluctuations are very small and do not depend significantly on the lattice spacing. One concludes 
that the reweighting of the rational approximation works very well.

\subsection{Physical observables}

\begin{figure}
\begin{center}
\includegraphics[width=0.49\textwidth]{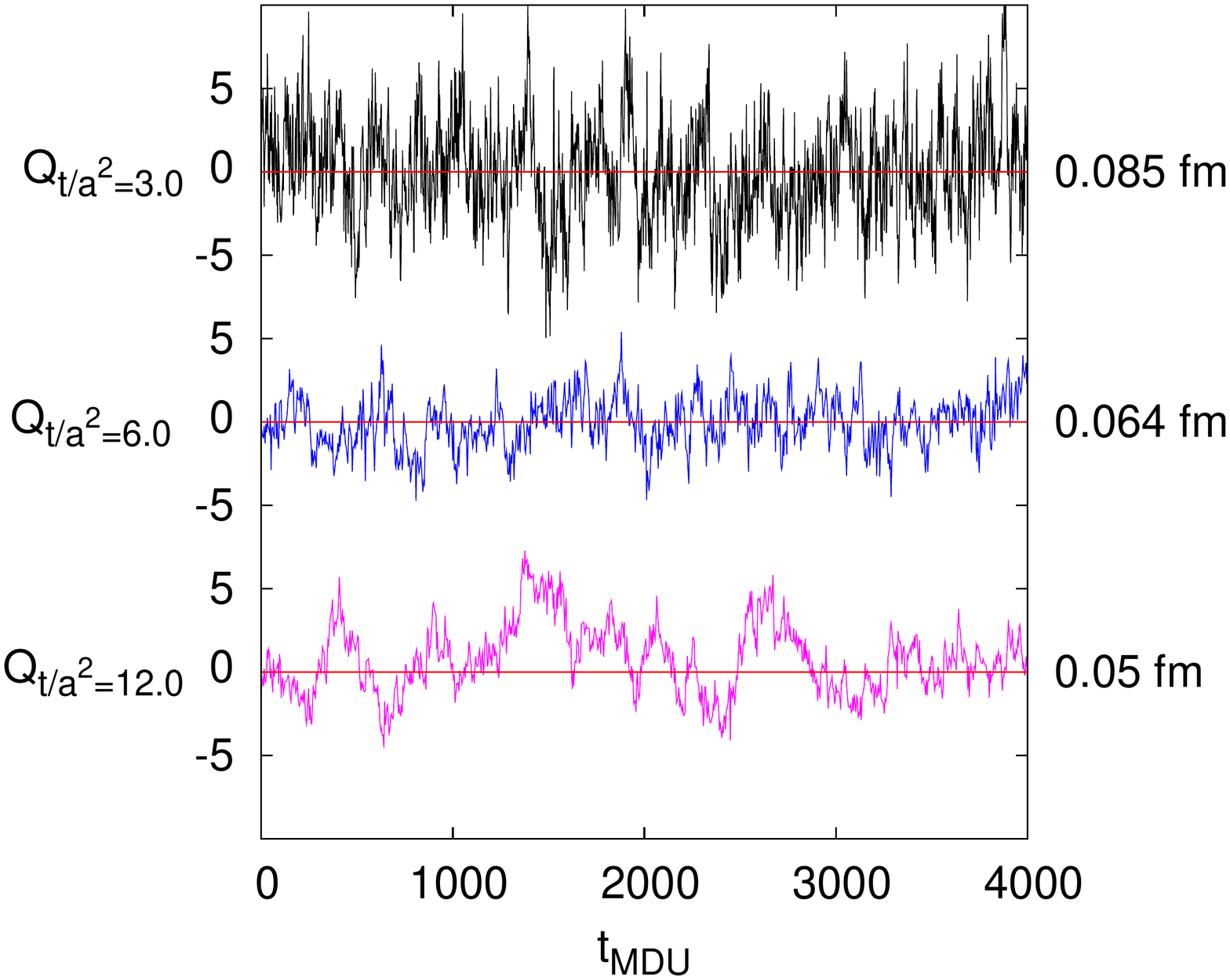}
\includegraphics[width=0.49\textwidth]{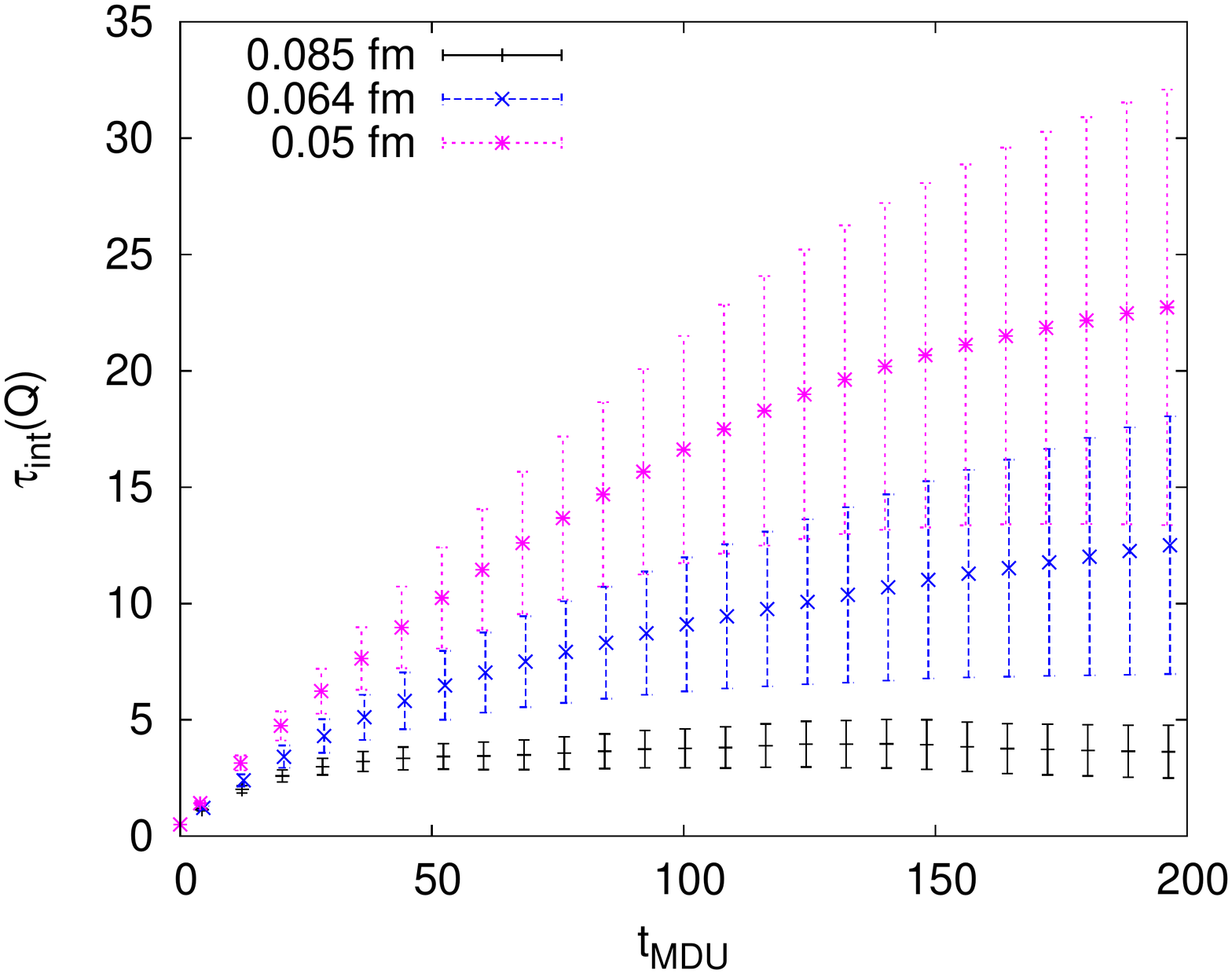}
\caption{Topological charge at the symmetric point at flow times $t/a^2={\color{black}3.0}, {\color{blue}6.0}, {\color{pink}12.0}$ which roughly reflect the scaling of $t_0$ with $a$. 
Data comes from ensembles H101r001, H200r000 and N300r002, see Ref.\cite{2+1}. \label{fig. topological charge}}
\end{center}
\end{figure}

Once the reweighting factors are computed one can evaluate expectation values of physical observables. 
First, we compute the topological charge given by the field-theoretic definition at Wilson flow time of the order of $t_0$. 
The left panel of figure \ref{fig. topological charge} shows the comparison of the history of the topological charge at three 
different lattice spacings evaluated on ensembles at the symmetric point. Similar information is plotted
on the right panel of that figure, where the integrated autocorrelation times are shown. One clearly sees an increase of $\tau_{int}$ 
as $a$ is decreased with a scaling in agreement with \cite{luscher_schaefer_11}.

\begin{figure}
\begin{center}
\includegraphics[width=0.49\textwidth]{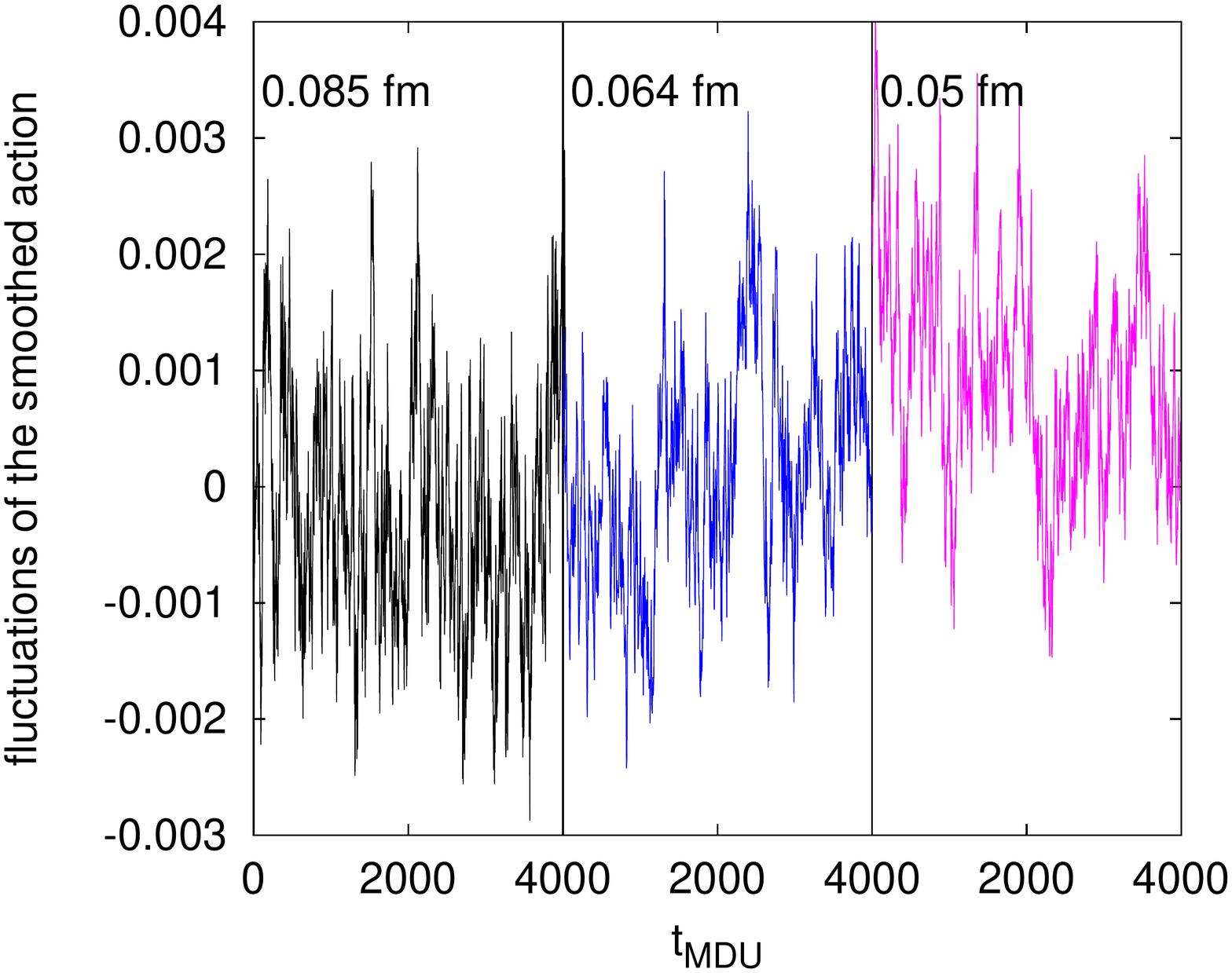}
\includegraphics[width=0.49\textwidth]{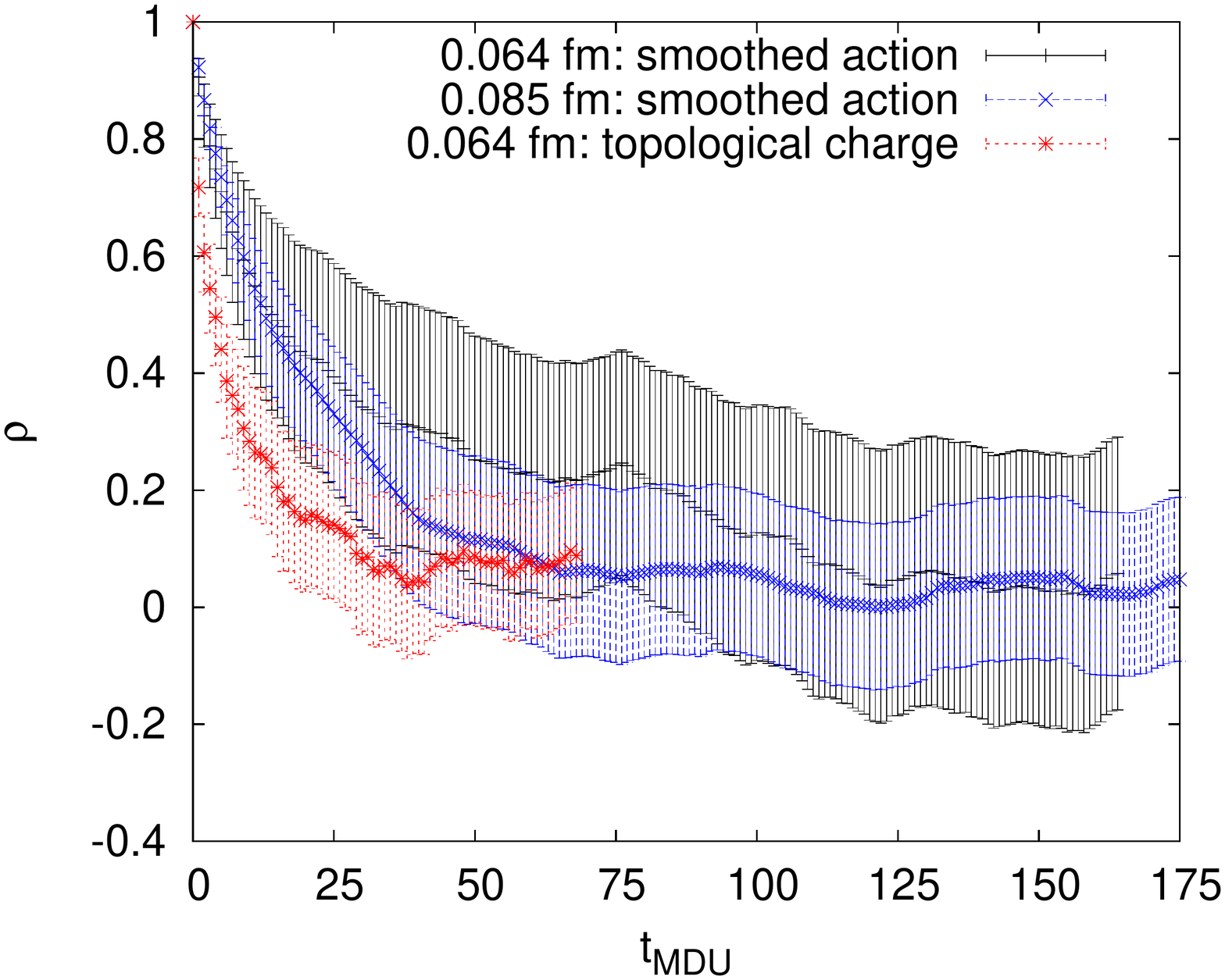}
\caption{The left panel shows the fluctuations of the YM action density at the symmetric point at 
flow time $t/a^2={\color{black}3.0}, {\color{blue}6.0}, {\color{pink}12.0}$. 
The right panel shows the comparison of autocorrelation functions of $E(t)$ and $Q(t)$.
Data comes from ensembles H101r000, H200r000 and N300r002 (only part of the full statistics is shown), 
see Ref.\cite{2+1}. \label{fig. action}}
\end{center}
\end{figure}

Next, we discuss the YM action density evaluated at Wilson flow time of the order of $t_0$. Figure \ref{fig. action} 
compares the MC history of the fluctuations of this observable for three different lattice spacings. Figure \ref{fig. action} also shows 
the autocorrelation functions of $E(t)$ corresponding to the two coarsest ensembles together with the autocorrelation function of
the topological charge. We notice that the integrated autocorrelation time of the topological charge is smaller, which is consistent
with the recent findings of \cite{bruno_14}.

\begin{figure}
\begin{center}
\includegraphics[width=0.49\textwidth]{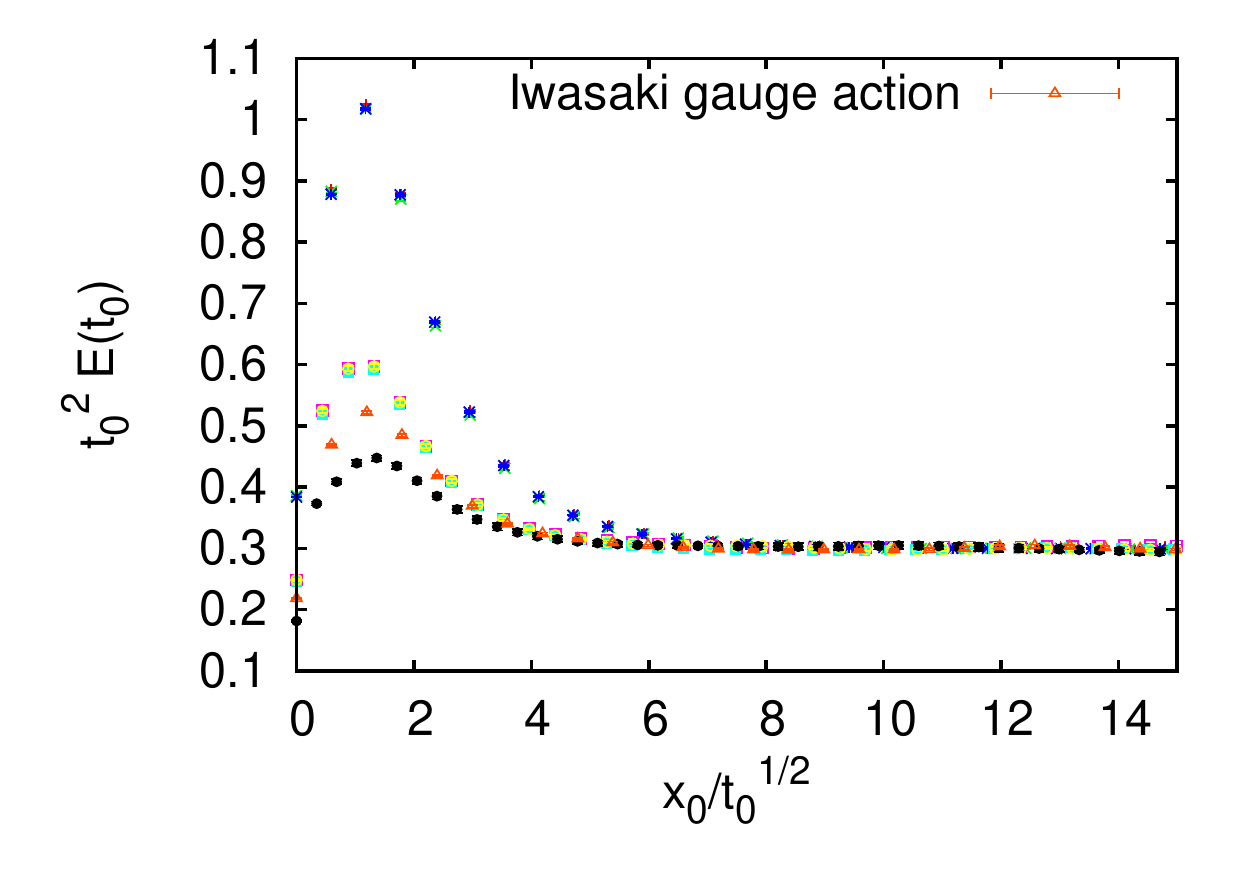}
\includegraphics[width=0.49\textwidth]{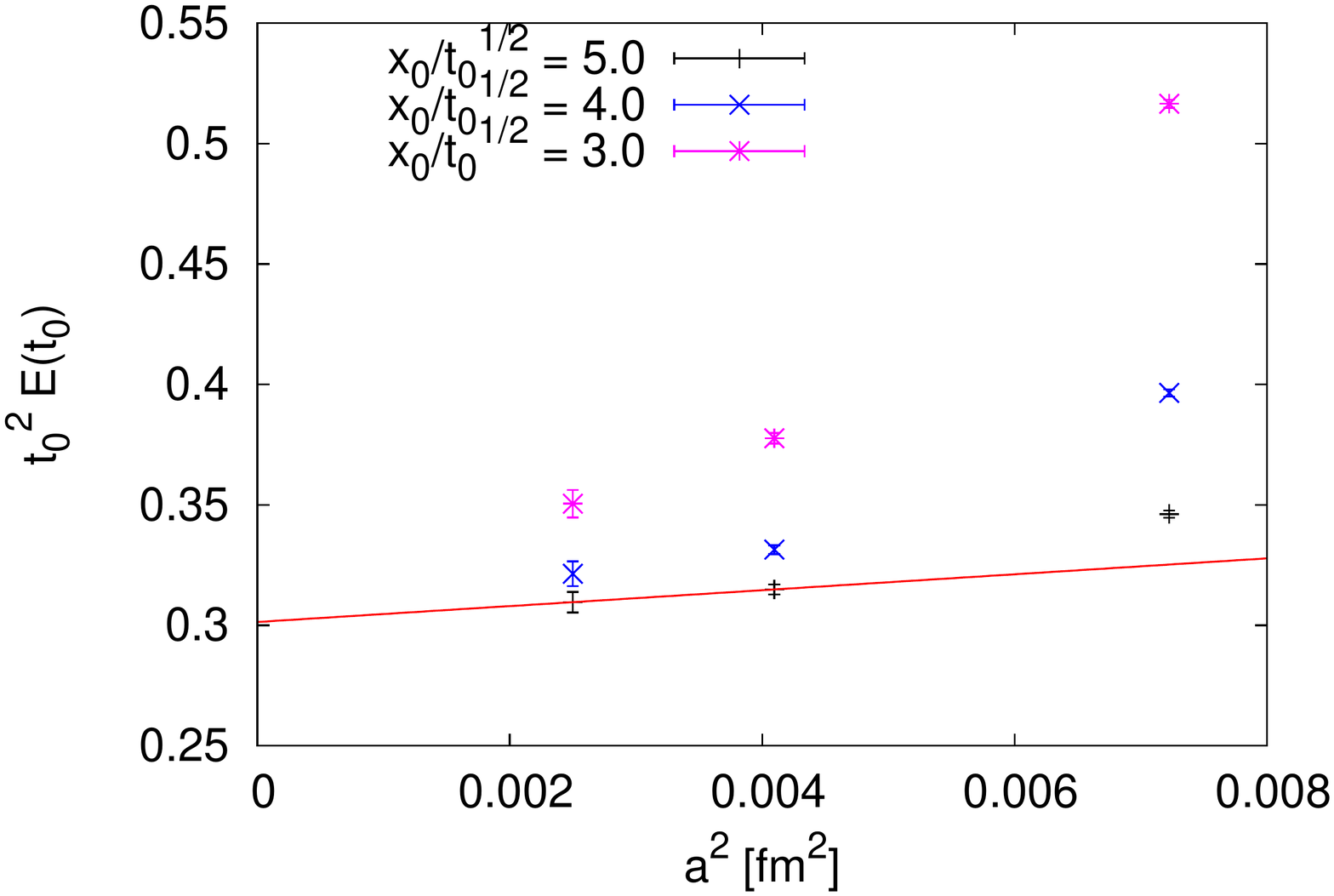}
\caption{$E(t_0,x_0)$ as a function of the distance from the boundary shows large cut-off effects. They exhibit strong lattice spacing 
dependence (see plot on the right), but are largely independent of the pion mass. \label{fig. cutoff}}
\end{center}
\end{figure}

$E(t)$ evaluated at each time-slice independently can be use to exhibit cut-off effects introduced by the open boundaries. 
On figure \ref{fig. cutoff} we plot $E(t)$ for different ensembles as a function of physical distance from the boundary 
expressed in terms of $x_0/\sqrt{t_0}$. The value approaches $0.3$ by difinition in the central region of the lattice. We see that 
the height of this lattice artifact depends strongly on the lattice spacings (the three distinctive sets of data points) and 
is roughly independent of the pion mass (different colors of the data points within each set). 
We also included results from a simulation with a Iwasaki gauge action at a lattice spacing of $a \approx 0.09$ fm. The discrepancy between 
these data and the data obtained with the Wilson action with a similar lattice spacing confirms the cut-off nature of the visible structure. On the right 
of figure \ref{fig. cutoff} we show the dependence of $E(t, x_0)$ on $a^2$ for three different values of $x_0/\sqrt{t_0}$. 
The data cannot be described by a linear dependence on the lattice spacing.

\begin{figure}
\begin{center}
\includegraphics[width=0.49\textwidth]{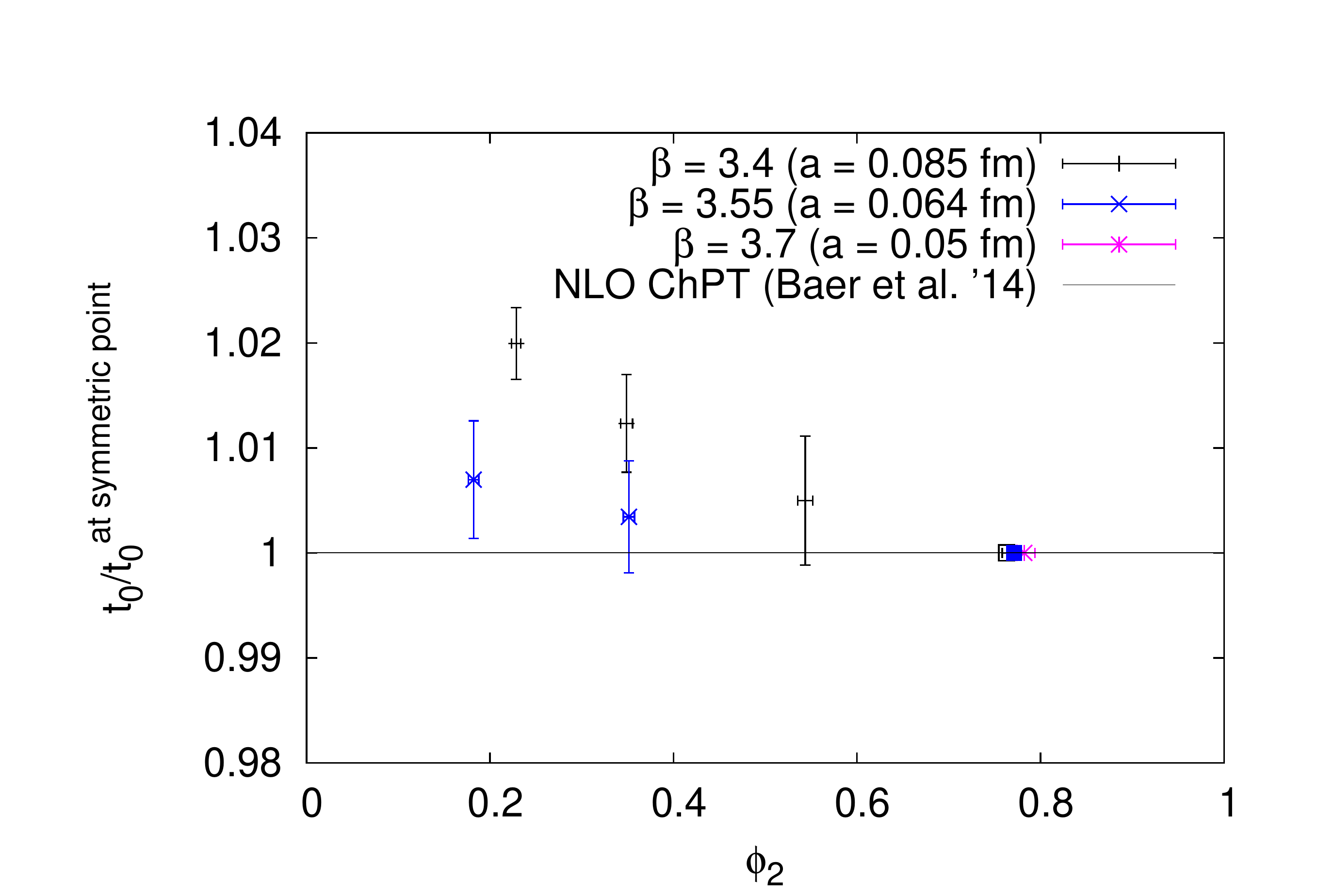}
\caption{Dependence of $t_0/t_0^{\textrm{ref}}$ on $m_{\pi}$ at different lattice spacings. 
The data show that the cut-off effects in this observable as well as its pion mass dependence are small, 
which is favorable in view of the tunning to the chiral trajectory.
\label{fig. t0}}
\end{center}
\end{figure}

Finally, we present our estimates of $t_0$ for all the considered ensembles normalized by the value obtained at the
symmetric point as a function of $m_{\pi}$, see figure \ref{fig. t0}. We also include in the plot the analytic 
prediction from NLO chiral perturbation theory worked out in \cite{golterman_baer_14}. We notice that the data points
deviate from this prediction by very small amounts; cut-off effects are small.

\section{Conclusions}

Several points should be highlighted. The algorithmic improvements turned out to work very well in practice. 
In all our runs the algorithm proved to be stable. The first preliminary measurements of physical observables 
show that the expected precision can be reached. The trajectory of constant trace of the mass matrix allowed 
precise matching and the tunning to the chiral trajectory turned out to be possible with small effort. 
The autocorrelation time of the topological charge is not the largest one, in agreement with Ref.\cite{bruno_14}.


\acknowledgments
The author would like to thank all the CLS collaboration and in particular M. Bruno, T. Korzec and S. Lottini for the help with the 
figures and S. Schaefer, R. Sommer and H. Simma for many useful discussions.


\begin{thebibliography}{99}
\bibitem{2+1} M. Bruno \emph{et al.}, arXiv:1411.3982
\bibitem{luscher_schaefer_11} M. L\"uscher, S. Schaefer,  JHEP 1107 (2011) 036, arXiv:1105.4749
\bibitem{luscher_palombi_08} M. L\"uscher, F. Palombi,  PoS LATTICE2008 (2008) 049, arXiv:0810.0946
\bibitem{luscher_04} M. Luscher, Comput. Phys. Commun. 156 (2004) 209-220, arXiv:hep-lat/0310048v1
\bibitem{openQCD} http://luscher.web.cern.ch/luscher/openQCD/
\bibitem{cls} Coordinated Lattice Simulations, https://wiki-zeuthen.desy.de/CLS/CLS
\bibitem{bulava_schaefer_13} J. Bulava, S. Schaefer, Nucl.Phys. B874 (2013) 188-197, arXiv:1304.7093
\bibitem{luscher_10} M. L\"uscher, PoS LATTICE2010 (2010) 015, arXiv:1009.5877
\bibitem{luscher_schaefer_13} M. L\"uscher, S. Schaefer, Comput.Phys.Commun. 184 (2013) 519-528, arXiv:1206.2809
\bibitem{borsanyi_12} S. Borsanyi \emph{et al.}, JHEP 1209 (2012) 010, arXiv:1203.4469
\bibitem{bruno_14} M. Bruno \emph{et al.}, JHEP 1408 (2014) 150, arXiv:1406.5363
\bibitem{mattia} M. Bruno \emph{et al.}, arXiv:1411.5207, in these proceedings
\bibitem{golterman_baer_14} M. Golterman, O. B\"ar,  Phys.Rev. D89 (2014) 034505, arXiv:1312.4999
\end{thebibliography}
\end{document}